\documentclass[useAMS,usenatbib,usegraphicx]{mn2e}

\usepackage{definitions}
\usepackage[T1]{fontenc}
\usepackage{mathptmx}
\usepackage{rotating}
\usepackage{amsfonts}
\usepackage{graphicx}
\usepackage{amsbsy}
\usepackage[fleqn]{amsmath}
\usepackage{xcolor}

\renewcommand{\v}[1]{\textbf{\textit{#1}}}

\let\mr=\mathrm

\newcommand{\dd}{\, \mathrm{d}}

\newcommand{\bq}{\begin{equation}}
\newcommand{\eq}{\end{equation}}

\newcommand{\C}{\boldsymbol{C}}

\newcommand{\vtheta}{\boldsymbol{\theta}}

\newcommand{\vpsi}{\boldsymbol{\psi}}
\newcommand{\vomega}{\boldsymbol{\omega}}

\newcommand{\LCDM}{\Lambda\mathrm{CDM}}

\title[Dark energy model selection]{Dark energy model selection with current and future data}
\author[I. Debono]{Ivan~Debono\thanks{E-mail: mail@idebono.eu}\\
{APC, AstroParticule et Cosmologie, Universit\'{e} Paris Diderot, CNRS/IN2P3, CEA/lrfu, Observatoire de Paris, Sorbonne Paris Cit\'{e}}\\{10, rue Alice Domon et L\'{e}onie Duquet, 75205 Paris Cedex
13, France.} \\{LESIA, Observatoire de Paris, CNRS, UPMC, Universit\'{e} Paris Diderot,}\\{5 place Jules Janssen, 92195 Meudon, France.}}
\begin{document}


\pagerange{\pageref{firstpage}--\pageref{lastpage}} \pubyear{2014}

\maketitle
\label{firstpage}

\begin{abstract}
The main goal of the next generation of weak-lensing probes is to constrain cosmological parameters by measuring the mass distribution and geometry of the low redshift Universe and thus to test the concordance model of cosmology. A future all-sky tomographic cosmic shear survey with design properties similar to \textit{Euclid} has the potential to provide the statistical accuracy required to distinguish between different dark energy models. In order to assess the model selection capability of such a probe, we consider the dark energy equation-of-state parameter $w_0$. We forecast the Bayes factor of future observations, in the light of current information from \textit{Planck}, by computing the predictive posterior odds distribution. We find that \textit{Euclid} is unlikely to overturn current model selection results, and that the future data are likely to be compatible with a cosmological constant model. This result holds for a wide range of priors.
\end{abstract}

\begin{keywords}
cosmology: dark energy -- weak gravitational lensing -- methods: statistical
\end{keywords}

\section{Introduction}
\label{Intro}

There is no shortage of data in modern cosmology, and information from various experiments has allowed us to measure the parameters in our cosmological model with increasing precision. These data include cosmic microwave background measurements (e.g. \citealt{WMAP9}, \textit{WMAP};  \citealt{Planck-Collaboration:2013aa}, \textit{Planck}), supernovae compilations (e.g. \citealt{Goldhaber:2009aa}, SCP), large scale structure maps (e.g. \citealt{Ahn2014}, SDSS), and weak-lensing observations (e.g. \citealt{Parker2007}; \citealt{Schrabback:2009}). The next generation of experiments (e.g. \citealt{Amendola2013}, \textit{Euclid}; \citealt{Blake2004}, SKA) should provide even better precision.

The $\Lambda$ cold dark matter ($\LCDM$) concordance model can fit current astrophysical data with only six parameters describing the mass-energy content of the Universe (baryons, CDM and dark energy) and the initial conditions. However, this is not a statement on whether the model is correct. It merely implies that deviations from $\LCDM$ are too small compared to the current observational uncertainties to be inferred from cosmological data alone. One obvious example is the addition of hot dark matter (HDM) to the model, i.e. parameters for the physical neutrino density and the number of massive neutrino species \citep[see e.g.][]{Abazajian:2011aa,Audren:2013aa, Basse2014}. Although massive neutrinos are not required by current cosmological observations, neutrino oscillation observations have shown that neutrinos have a non-zero mass. 

The fundamental questions facing cosmologists are not simply about parameter estimation, but about the possibility of new physics, therefore about model selection. In addition to estimating the values of the parameters in the model, this involves decisions on which parameters to include or exclude. In some cases, the inclusion of parameters is possible only by invoking new physical models.

This is the case with the dark energy problem \citep{DETF, Peacock:2006aa}. There is firm observational evidence suggesting that the Universe entered a recent stage of accelerated expansion. The physical mechanism driving this expansion rate remains unclear and there exist several potential models. In the framework of General Relativity applied to a homogeneous and isotropic universe, the acceleration could be produced either by an additional term in the gravitational field equations or by a new isotropic comoving perfect fluid with negative pressure, called dark energy \citep[see e.g.][]{Peebles:2003,Johri:2007,Polarski:2013aa}. 

The main science goal of the next generation of cosmological probes is to test the concordance model of cosmology. In the case of dark energy, the objective is to measure the expansion history of the Universe and the growth of structure (see \citealt{Weinberg2013} for a comprehensive review). In this work, we will consider cosmic shear from a future all-sky survey similar to the European Space Agency mission \textit{Euclid}, due for launch in 2020 \citep[see][]{Amendola2013}. The main scientific objective of \textit{Euclid} is to understand the origin of the accelerated expansion of the Universe by probing the nature of dark energy using weak-lensing and galaxy-clustering observations. It could potentially test for departures from the current concordance model \citep[see e.g.][]{Heavens:2007,Jain:2010aa,Zhao:2012aa}. In a previous paper \citep{Debono:2013} we studied the ability of \textit{Euclid} weak lensing to distinguish between dark energy models. This time we go a step further by addressing the following question: Based on current data, what model will we select using future weak-lensing data from \textit{Euclid}? What is the probability that these data could favour $\Lambda$CDM?

Cosmologists are faced with the task of constructing valid physical models based on incomplete information. In this relation between data and theory, Bayesian inference provides a quantitative framework for plausible conclusions (see e.g. \citealt{Robert2008,Hobson:2010aa,Jenkins:2011aa} for a discussion) and can be understood as operating on three levels:
\begin{enumerate}
\item Parameter inference (estimation): we assume that a model $M$ is true, and we select a prior for the parameters $p(\vtheta | M)$, where $p$ is some probability distribution and $\vtheta$ is the set of model parameters.
\item Model selection or comparison: there are several possible models $M_i$. We find the relative plausibility of each in the light of the data $D$.
\item Model averaging: there is no clear evidence for a best model. We find the inference on the parameters which accounts for the model uncertainty.
\end{enumerate}
The dark energy question is a model comparison or model averaging problem. In this work, we will confine ourselves to model selection. 

In order to produce an accelerated expansion at the present epoch, the dark energy equation-of-state parameter should satisfy the conservative bound $w_\mr{DE}={p_\mr{DE} }/{\rho_\mr{DE}}< -0.5$. Observations suggest a lower value, close to $-1$. If the data are compatible with this value, then in model selection terms it means they are compatible with $\LCDM$. However, $\LCDM$ is not merely a special case of some more general model where $w_\mr{DE}=-1$. It contains a smaller number of free parameters, and if it fits the data, it is favoured by the Occam razor effect because it is more predictive. In the cosmological context, the question is whether there is evidence that we need to expand our cosmological model beyond $\Lambda$CDM to fit these data \citep[see e.g.][]{Liddle:2004aa,March:2011aa}.

Consider a result from an experiment quoted in terms of a mean value $\mu$, and a confidence interval $\sigma$. This is a parameter estimation result. In the frequentist interpretation, a confidence interval of $68.3$ per cent means that if we were to repeat the experiment an infinite number of times and obtain a $1\sigma$ distribution for the mean, the true value would lie inside the intervals thus obtained $68.3$ per cent of the time. This is not the same as saying that the probability of $\mu$ lying within a given interval is $68.3$ per cent. The latter is a statement on model selection, and it only follows if we use Bayesian techniques.

In simple terms, how do we know if a given accuracy on a certain mean value enough to falsify a model? If it falsifies a model, what does it verify? It is therefore clear that model selection calculations must include information on the alternatives under consideration. We cannot reject a hypothesis unless an alternative hypothesis is available that fits the facts better.

In the context of the dark energy problem, it means that a claim such as `$\Lambda$CDM is false' is not enough. We need an alternative model, for which we would know at least the number of free parameters and their allowed ranges, before the data come along. This is the model prior. Intuitively, we know that the prior affects the model selection outcome. We know $w_\mr{DE}$ to be within  range of values around $-1$, but for a cosmological constant, the prior width is zero. Statistically, we will measure $w_\mr{DE}$ to be a different value each time, so we take some average. It is the average that is compatible with theory that we understand to be the value of the cosmological constant.

The question is therefore to know which range of measured values leads us to choose $\Lambda$CDM, and which values lead us to discard it. In terms of model selection, we need to quantify the degree of compatibility  with $\Lambda$CDM of a measured value of $w_\mr{DE}$ that is $x \sigma$ away from $-1$.

One important point to note is that in constructing models, we are seeking to find that model which will best predict future data. We can always include all possible parameters and obtain a perfect fit to the current data, but we also want our model to be predictive. Thus, the model that explains the past data best may not be most predictive model. Bayesian evidence quantifies this trade-off between goodness of fit and predictivity or model simplicity.

We are interested in forecasting the result of a future model comparison, by predicting the distribution of future data. In this work, we assess the potential of \textit{Euclid} to address model comparison questions, based on current information. We derive a predictive distribution for the dark energy equation-of-state parameter for a cosmic shear survey with the \textit{Euclid} probe using the predictive posterior odds distribution (PPOD) method developed by \citet{Trotta:2007aa}. We also study the dependence of our results on the prior width. 

This paper is organized as follows. In Section \ref{Bayesian_inference}, we describe the Bayesian framework. The PPOD method is described in Section \ref{PPOD_section}. Our cosmological and weak-lensing formalism for \emph{Euclid}, together with the current and future data are described in Section \ref{Method}. We apply the PPOD to \textit{Euclid} in Section \ref{Section_Results}, and present our conclusions in Section \ref{Conclusions}.

\section{Bayesian model comparison}
\label{Bayesian_inference}

The details of Bayesian model comparison and the derivation of the Savage-Dickey density ratio (SDDR) are given in a companion paper \citep{Debono:2013}. Here we give a brief overview.

Bayesian inference is based on the logic of probability theory. The product rule gives us Bayes' theorem:
\bq\label{eqn1}
p(\vtheta | d,M)=\frac{p(d | \vtheta,M)p (\vtheta | M)}{p( d | M)}.
\eq
The left-hand side is the posterior probability for the vector of unknown model parameters $\vtheta$ of length $n$ given the data $d$ under model $M$. The prior probability distribution function $p(\vtheta|M)$ is an expression of our state of knowledge before observing the data. This defines the prior available parameter space under the model $M$. The denominator $p (d | M)$ is the Bayesian evidence or model likelihood. This is the probability of observing the data $d$ given that the model $M$ is correct. 

Model selection usually involves the calculation of the evidence. This may be expressed as the multidimensional integral of the likelihood over the prior, or the parameter space under $M$:
\bq
p (d | M) = \int  p(d | \vtheta,M) p(\vtheta | M) \dd\vtheta ,
\eq
where $\vtheta$ is in general multidimensional and $d$ is a collection of measurements (current or future).

The Bayes factor is then the ratio of the evidence for two competing models $M_0$ and $M_1$, or the ratio of posterior odds:
\bq
B_{01}\equiv\frac{p(d | M_0)} {p(d | M_1)}=\frac{p(M_0|d)}{p(M_1|d)}.
\eq

In many applications, such as the dark energy problem, the models are nested within each other. Let us write the vector of model parameters as $\vtheta=({\boldsymbol{\psi},\vomega})$. The model $M_0$, containing the vector of free parameters $\boldsymbol{\psi}$ is a restricted submodel of $M_1$, which contains the parameters $\boldsymbol{\psi}$ and $\boldsymbol{\omega}$. In $M_0$, the additional parameters are fixed at $\boldsymbol{\omega}=\boldsymbol{\omega}_\ast$. In the $\Lambda$CDM model, the dark energy equation-of-state parameter $w_\mr{DE}$ is fixed at $-1$. We assume separable priors, i.e.
\bq
p(\boldsymbol{\omega},\boldsymbol{\psi}|M_1)=p(\boldsymbol{\omega}|M_1)p(\boldsymbol{\psi}|M_0).
\eq

Then, the Bayes factor can be written as the ratio of the marginalized posterior over the prior marginal density of $\vomega$ under the extended model $M_1$, evaluated at the value $\vomega=\vomega_\ast$: 
\bq \label{SDDR}
B_{01}=\frac{p(\vomega|d,M_1)}{p(\vomega|M_1)}\bigg|_{\vomega=\vomega_\ast},
\eq
which is the SDDR \citep{Dickey1971}. The SDDR expresses the Bayes factor as an amount of information brought by the data. It is therefore a good tool for model selection in cosmology.

We use the Jeffreys scale to interpret the logarithm of the Bayes factor in terms of the strength of evidence. We adopt a slightly more conservative version of the convention used by \citet{Jeffreys:1961} and \citet{Trotta:2007}. This is shown in Table \ref{Jeffreys}.

\begin{table}
\caption{Jeffreys's scale for the strength of evidence when comparing two models $M_0$ (restricted) against $M_1$ (extended), interpreted here as the evidence for the extended model. The probability is the posterior probability of the favoured model, assuming non-committal priors on the two models, and assuming that the two models fill all the model space. Negative evidence for the extended model is equivalent to evidence for the simpler model. Note that the labels attached to the Jeffreys scale are empirical, and their interpretation depends to a large extent on the problem being modelled. An experiment for which $|\ln B|<1$ is usually deemed inconclusive. }
\begin{center}
\begin{tabular}{@{} lll @{}}
\hline
$\ln B_{01} $ 		 		& Probability 	&Evidence\\
\hline
$>0$	  	& $<0.5$ 	 & Negative\\
$-2.5$ to $0$ 		& $0.5$ to $0.923$ 	& Positive \\
$-5$ to $-2.5$ 	 		& $0.923$ to $0.993$ 	&  Moderate \\
$< -5$ 			& $>0.993$ 	 &  Strong \\
\hline
\end{tabular}
\end{center}
\label{Jeffreys}
\end{table}

Inconclusive evidence for one model means that the alternative model cannot be distinguished from the null hypothesis. This occurs when $|\ln B_{01}|<1$. A positive Bayes factor $\ln B_{01}>0$ favours model $M_0$ over $M_1$ with odds of $B_{01}$ against 1.

\section{The PPOD}
\label{PPOD_section}

In a companion paper \citep{Debono:2013}, we examined the ability of \textit{Euclid} cosmic shear measurements to distinguish between different dark energy models. Our current knowledge is included in the calculations in two ways: through our choice of fiducial model, and through the prior ranges. Implicit in the former is the assumption that the future maximum likelihoods for the cosmological parameters common to all models under consideration will be roughly the same as the current likelihoods. However, we do not include information on the position of the current maximum likelihood of the extra parameters (in this case, the dark energy equation-of-state parameters). In other words, we do not take into account the present posterior distribution.

One way of including this information is through the PPOD developed by \citet{Trotta:2007aa}. This extends the idea of posterior odds forecasting introduced by \citet{Trotta:2007} and also \citet{Pahud:2006aa,Pahud:2007aa}, which is based on the concept of predictive probability. This uses present knowledge and uncertainty to predict what a future measurement will find, with corresponding probability. The predictive probability is therefore the future likelihood weighted by the present posterior.

The PPOD is a hybrid technique, combining a Fisher matrix analysis \citep{Fisher1935,Fisher1936} with the SDDR. It gives us the probability distribution for the model comparison result of a future measurement. It is conditional on our present knowledge, and gives us the probability distribution for the Bayes factor of a future observation. In other words, it allows us to quantify the probability with which a future experiment will be able to confirm or reject the null hypothesis.

The PPOD showed its usefulness in predicting the outcome of the \textit{Planck} experiment. \citet{Trotta:2007aa} found that \textit{Planck} had over 90 per cent probability of obtaining model selection result favouring a scale-dependent primordial power spectrum, with only a small probability that it would find evidence in favour of a scale-invariant spectrum. This result was confirmed by actual data a few years later, when \textit{Planck} temperature anisotropy measurements combined with the \textit{Wilkinson Microwave Anisotropy Probe (WMAP)} large-angle polarization found a value of $n_\rmn{s}=0.96\pm 0.0073$, ruling out scale invariance at over $5\sigma$ \citep{Planck-Collaboration:2013ab}.

In this paper, we apply the PPOD technique to the dark energy equation-of-state parameter, comparing the evidence for $\Lambda$CDM against a dynamical dark energy model $w$CDM. From the current posterior, we can produce a PPOD for the \textit{Euclid} satellite. In this section we review the formalism of the PPOD.

The predictive distribution for future data $D$ is
\begin{align}
p(D|d)&=\sum_{i=0}^1p(D|d,M_i)p(M_i|d)\nonumber\\
&=\sum_{i=0}^1p(M_i|d)\int p(D|\vtheta,M_i)p(\vtheta |d,M_i)\dd \vtheta,
\end{align}
where $d$ is the current data, and the sum runs over the two models we are considering. In the equation above, $p(D|\vtheta,M_i)$ is the predicted likelihood for future data, assuming $\vtheta$ is the correct value for cosmological parameters under model $M_i$. We obtain a Gaussian approximation to the future likelihood by performing a Fisher matrix analysis assuming $\vtheta$ as a fiducial model. This gives us a forecast of the parameter covariance matrix \textbf{\textsf{C}} for future data $D$.

The PPOD for the future Bayes factor $B_{01}$, conditional on current data $d$ is then 
\begin{align}
p(B_{01} | d) &=\int p(B_{01}, D | d) \rmn{d}D \nonumber\\
& =\int p(B_{01} | D , d)p(D|d)\rmn{d} D.
\end{align}
We can calculate the Bayes factor as a function of the future data only, i.e. $B_{01}(D)$, from equation (\ref{eqn10}). Using a property of the Dirac delta-function we can then express the PPOD as \citep{Trotta:2007aa}:
\bq
p(B_{01} | d)=\int\delta(D-B_{01}(D))p(D|d)\rmn{d}D,
\eq
where $\delta$ is the Dirac delta-function.

Let us consider the case of nested models, with ${\vtheta=(\vpsi,\vomega)}$ as defined previously. It is reasonable to assume that the current and future likelihoods for the data considered in this paper are both Gaussian. For future data, this assumption is implicit in our use of Fisher matrix analysis to forecast the future covariance matrix \textbf{\textsf{C}}. We make the further assumption that the covariance matrix does not depend on the fiducial values chosen for the common parameters $\vpsi$. Then, we can marginalize over the common parameters, and compare a one-dimensional model $M_1$ with a model $M_0$ with no free parameters.

The priors on the extra parameter are taken to be Gaussian, centred on zero, with a prior width equal to unity. The current likelihood is also assumed to be Gaussian, centred on $\omega=\mu$ of width $\sigma$. The Gaussian mean and width are expressed in units of the prior width and are therefore dimensionless. Likewise, the predicted likelihood is assumed to have a Gaussian distribution, with mean $\omega=\nu$ and constant standard deviation $\tau$. The latter is the forecast error $\tau=\sqrt{\C_{11}}$ obtained from a Fisher matrix calculation. It is assumed to be independent of $\omega$, which is a reasonable assumption, since the marginalised errors are very stable to a change in the fiducial values of the model parameters over the region of interest (see \citealt{Debono:2013} for the variation of the dark energy figure-of-merit in the $w_0-w_a$ parameter space). 

The predictive distribution can then be expressed analytically as
\begin{align}
p(D|d)\propto &\frac{p(M_0)}{\tau\sigma}\exp \left(-\frac{1}{2}\frac{\nu^2\sigma^2+\mu^2\tau^2}{\tau^2\sigma^2}\right)+\nonumber\\
&\frac{p(M_1)}{\sqrt{\tau^2+\sigma^2+\tau^2\sigma^2}}\exp\left(-\frac{1}{2}\frac{(\nu-\mu)^2+\sigma^2\nu^2+\tau^2\mu^2}{\tau^2+\sigma^2+\tau^2\sigma^2}    \right) 
\end{align}
where the normalizing constants for the probability distribution are left out. This gives the probability of obtaining a value $\omega=\nu$ from a future measurement as a function of the future mean $\nu$ conditional on the present data $d$.

Note that in the equation above, $p(M_0)=p(M_1)=\tfrac{1}{2}$. At present, there is no evidence which justifies assigning a higher probability to a particular model. This is a statement on our prior knowledge, which is based on the accumulation of information from a multitude of experiments \citep[see][]{Brewer:2009aa}. In this paper, we justify assigning equal probabilities to each model because we are testing two at a time. 

The PPOD is obtained by applying the SDDR using the relation between $\nu$ and the future model selection outcome $B_{01}(D)$ \citep{Trotta:2007aa}:
\bq \label{eqn10}
\nu^2=\tau^2(1+\tau^2)\left(\ln\frac{1+\tau^2}{2\upi \tau^2}-2\ln B_{01}(D)  \right).
\eq
This relation only holds for a Gaussian prior and a posterior distribution that is accurately described by a Gaussian. The latter assumption will most likely not hold in the tails of the distribution, where $|\nu-\omega_\ast|/\tau\gg1$. In other words, this is the region where the mean value of the extra parameter in the extended model is many sigmas away from its fixed value in the restricted model.  In this case, parameter estimation should be enough to provide evidence against $M_0$, even though we might not be able to calculate a precise value for the expected odds.

\section{Forecasts for \textit{Euclid}}
\label{Method}

We apply the PPOD technique to assess the potential of the \textit{Euclid} mission in terms of model selection, taking into account the information from current data. Our current information is taken from \textit{Planck} results, while we forecast our future cosmic shear data from \textit{Euclid}.

\subsection{The future data}

We forecast the errors for future \textit{Euclid} cosmic shear data using the Fisher matrix technique. The restricted fiducial cosmological model used for our forecast contains parameters describing baryonic matter, CDM, massive neutrinos (or HDM) and dark energy. We drop the requirement for flat spatial geometry by including a dark energy density parameter $\Omega_\mr{DE}$ together with the total matter density $\Omega_\rmn{m}$. 

We choose fiducial parameter values based on the \textit{Planck} 2013 best-fitting values \citep{Planck-Collaboration:2013aa}:
\begin{enumerate}
\item Total matter density: $\Omega_\rmn{m}=0.31$ (which includes baryonic matter, HDM and CDM), 
\item Baryonic matter density: $\Omega_\rmn{b}=0.048 $,
\item Neutrinos (HDM): $m_\nu=0.25\,\mr{eV}$ (total mass); ${N_\nu=3}$ (number of  massive neutrino species),
\item Dark energy density: $\Omega_{\mathrm{DE}}=0.69$,
\item Hubble parameter: $h=0.67(100\,\mr{km}^{-1}\mr{Mpc}^{-1})$ and
\item Primordial power spectrum parameters: $\sigma_8=0.82$ (amplitude); $n_\rmn{s}=0.9603$ (scalar spectral index); $\alpha=0$ (its running).
\end{enumerate}

We assume a total of three neutrino species, with degenerate masses for the most massive eigenstates. The temperature of the relativistic neutrinos is assumed to be equal to $(4/11)^{1/3}$ of the photon temperature. We model $N_\nu$, the number of massive neutrino species, by a continuous variable. 

CMB anisotropy observations from the \textit{Planck} probe suggest caution in employing an overly simple parametrization of the primordial power spectrum \citep{Planck-Collaboration:2013aa,Planck-Collaboration:2013ab}. For this reason, we allow for possible departures from a scale-invariant primordial power spectrum.

For simplicity, we shall refer to this fiducial model as $\Lambda$CDM. Note that our work implicitly assumes that future best-fit values for the restricted model will not deviate significantly from the current ones. This assumption must be used carefully (see \citealt{Starkman:2010aa}) but it is a reasonable one when studying the question of nested models. In this case, the model doubt really only concerns the need for the additional parameters in the extended model (see \citealt{Starkman:2008aa,March:2011aa}). 

We use the numerical Boltzmann code \textsc{camb} \citep{CAMB} to calculate the linear matter power spectrum. This includes the contribution of baryonic matter, cold dark matter, dark energy and massive neutrino oscillations. We use the \citet{Smith:2003aa} \textsc{halofit} fitting formula to calculate the non-linear power spectrum, with the modification by \citet{Bird2012}. The power spectrum is normalized using $\sigma_8$, the root mean square amplitude of the density contrast inside an $8\,h^{-1}\mr{Mpc}$ sphere. 

To calculate future errors, we use forecasts for an all-sky tomographic weak-lensing survey similar to \textit{Euclid} \citep{Amendola2013,Laureijs:2011aa}, using the method described in \citet{Debono:2013}. 

We follow the power spectrum tomography formalism in \citet{Hu:2004}, first proposed by \citet{Hu1999a}, with the background lensed galaxies divided into 10 redshift bins. Cross-correlations of shears are carried out within and between bins. The 3D power spectrum is projected onto a 2D lensing correlation function using the \citet{Limber:1953} equation
\bq C_\ell^{ij}=\int \mr{d}z \frac{H}{D^2_A} W_i(z)W_j(z)P(k=\ell/D_A,z),\eq
where $i$, $j$ denote different redshift bins. The weighting function $W_i(z)$ is defined by the lensing efficiency: 
 \bq  
W_i(z)=\frac{3}{2}\Omega_\rmn{m} \frac{H_0}{H}\frac{H_0 D_\rmn{OL}}{a}\int_z^\infty \mr{d} z^{\prime}\frac{D_\rmn{LS}}{D_\rmn{OS}}P(z^{\prime}),
\eq where the angular diameter distance to the lens is $D_\rmn{OL}$, the distance to the source is $D_\rmn{OS}$, and the distance between the source and the lens is $D_\rmn{LS}$ (see \citealt{Hu:2004}). Our multipole range is $10<\ell<5000$. Due to the Limber approximation, there is a correspondence between the spatial wavenumber $k$ and the angular wavenumber $\ell$. In order to cover the chosen multipole range for the survey's median redshift, we use a maximum value of $10^3$ for $k$ in our calculations. We assume the \citet*{Smail1994} probability distribution function for the galaxies
\bq P(z)=z^a \exp\left[-\left(\frac{z}{z_0}\right)^b\right],\eq where $a=2$ and $b=1.5$, and $z_0$ is determined by the median redshift of the survey $z_\mr{median}$. 

We calculate the measurement errors based on two configurations of the \textit{Euclid}-type survey, referred to as the `requirements' and `goals' in the \textit{Euclid} Definition Study Report \citep{Laureijs:2011aa}. The experiment is defined by the following parameters: the survey area $A_\mathrm{s}$, median redshift of the density distribution of  galaxies $z_\mr{median}$, the observed number density of galaxies $n_\mathrm{g}$, the photometric redshift errors $\sigma_z(z)$ and the intrinsic noise in the observed ellipticity of galaxies $\sigma_\epsilon$, such that $\sigma_\epsilon^2=\sigma_\gamma^2$, where $\sigma_\gamma$ is the variance in the shear per galaxy. These parameters are shown in Table \ref{survey_params}. 

\begin{table}
\begin{center}
\caption{Fiducial parameters for the \textit{Euclid}-type all-sky weak-lensing survey used for our future data.}
\label{survey_params}
\begin{tabular}{@{}lrr@{}}
\hline
Survey property & Requirements & Goals\\
\hline
$A_\mathrm{s}$/sq degree & 15 000 &20 000\\
$z_\mr{median}$&0.9 & 0.9\\
$n_\mathrm{g}/\mr{arcmin}^{2}$&30 & 40\\
$\sigma_z(z)/(1+z)$& 0.05 &0.03\\
$\sigma_\epsilon$&0.25 & 0.25\\
\hline
\end{tabular}
\end{center}
\end{table}

The Fisher matrix for the shear power spectrum is given by \citep{Hu:2004}
\bq F_{\alpha\beta} = f_\mr{sky}\sum_\ell{\frac{(2\ell+1)\Delta \ell}{2}}\mr{Tr}\left[D_{\ell\alpha}\widetilde{C}_\ell^{-1}D_{\ell\beta}\widetilde{C}_\ell^{-1}\right],\eq
where the sum is over bands of multipole $\ell$ of width $\Delta \ell$, $\mr{Tr}$ is the trace, and $f_\mr{sky}$ is the fraction of sky covered by the survey. We assume the likelihood to have a Gaussian distribution, with zero mean. From the Fisher matrix we calculate the covariance matrix, which gives us the error forecasts on the parameters in our model.

\subsection{Dark energy parametrization}
\label{DE_section}

This paper examines the question of whether dark energy is $\Lambda$ or whether there is evidence for dynamical dark energy. Specifically, we ask how well the future \textit{Euclid} probe will be able to answer this in the light of the current model selection outcome.

Here, we have a case of a restricted model $\Lambda$CDM nested within an extended model, which we call $w$CDM. We consider an extension of $\Lambda$CDM by adding two dark energy parameters: the equation-of-state parameter at the present epoch $w_0$ and its variation $w_a$. The dynamical dark energy equation-of-state parameter, $w=p/\rho$, is expressed as function of redshift and is parametrized by  a first-order Taylor expansion in the scale factor $a$ \citep{ChevPol2001, Linder:2003}: \bq w(a)=w_0 + (1 -a)w_a,\eq where $a=(1+z)^{-1}$. This parametrization is motivated by the quintessence model, in which dark energy is some minimally coupled scalar field, slowly rolling down its potential such that it can have negative pressure. Scalar field models typically have a time-varying $w\geqslant -1$, and constant $w\neq-1$ models are poorly motivated, which is why we include the parameter $w_a$. In this study, however, we will focus on the value of $w_0$.

We include dark energy perturbations in all our calculations by using the parametrized post-Friedmann framework \citep{Hu:2007aa,Hu:2008aa} as implemented in \textsc{camb} \citep{Fang:2008aa,Fang:2008ab}.

As stated previously, we assume uncorrelated priors for the parameters in the restricted cosmological model and the extra parameter. The joint prior of $(w_0, \psi)$ is simply the product of the individual priors of $w_0$ and $\psi$. This allows us to use the SDDR (equation \ref{SDDR}) to find the Bayes factor. Strictly speaking, this is not the case ($\psi$ includes $w_a$, for instance), but our assumption is justified if we consider the form of the prior to be the space of possible choices, and not the space of actual observed data. In calculating the SDDR, we assume total ignorance of the possible values of the parameters in $\psi$. In other words, we take `ignorance' to be consistent with `independence'.

\subsection{The current data}
\label{The_current_data}

As our current posterior, we use the results from four \textit{Planck} data sets used by the \citet{Planck-Collaboration:2013aa} to estimate the values of cosmological parameters. In these parameter-estimation calculations, the \textit{Planck} temperature power spectrum is combined with a \textit{WMAP} polarization low-multipole likelihood and with four other data sets, as detailed below:
\begin{enumerate}
\item \textit{Planck}+WP+BAO: \textit{Planck} and \textit{WMAP}, combined with baryon acoustic oscillation measurements;
\item \textit{Planck}+WP+Union 2.1: \textit{Planck} and \textit{WMAP}, combined with an updated Union2.1 supernova compilation by \citet{Suzuki2012};
\item \textit{Planck}+WP+SNLS: \textit{Planck} and \textit{WMAP}, combined with the Supernova Legacy Survey compilation by \citet{Conley2011}; and
\item \textit{Planck}+WP+$H_0$: \textit{Planck} and \textit{WMAP}, combined with the \citet{Riess2011} $H_0$ measurements. \end{enumerate} 

We use a Gaussian approximation for the current likelihoods of the dark energy equation-of-state parameter. The values for the mean $w_0$ and width $\sigma$ for the likelihood from each dataset are given in Table \ref{Current_bayes}. From each current likelihood, using a Gaussian prior centred on $\omega_\ast$ with width $\Delta \omega$, we can calculate the Bayesian evidence using \citep{Trotta:2007}
\bq  \label{current_B}
\ln B_{01}(\beta,\lambda)=\frac{1}{2}\ln(1+\beta^{-2})-\frac{\lambda^2}{2(1+\beta^2)},
\eq
where $\lambda=|\mu-\omega_\ast |/\sigma$ and $\beta=\sigma/\Delta\omega$. We choose the prior width to be 0.5, and we shift and rescale the parameters so the Gaussian likelihood is centred on zero and parameters are dimensionless. Thus, $\lambda$ is the discrepancy between the mean value of $w_0$ and $w_0=-1$ expressed in number of sigmas. The quantity $\beta$ is the factor by which the prior accessible space is reduced by the data. It is evident from equation (\ref{current_B}) that Bayesian evidence is a function of both the data and the prior. 

\begin{table*}
\begin{minipage}{180mm}
 \caption{Model selection results with four current data sets including \textit{Planck}, assuming a Gaussian prior centred on $w_0=-1$ with a prior width of $0.5$. Most of the data favour $\Lambda$CDM. A combination of \textit{Planck}, WP and $H_0$ data shows positive evidence for dynamical dark energy. Note that this calculation uses a rather restrictive prior. This model of dynamical dark energy would have an equation-of-state parameter in the range $-1.5<w_0<-0.5$.}
  \label{Current_bayes}
\begin{center}
  \begin{tabular}{@{} lllll}
\hline
Current data & $w_0$ & $\sigma$ &$\ln B_{01}$ & Evidence\\ \hline
\textit{Planck}+WP+BAO & -1.13&0.120&0.900 & Positive for $\Lambda$CDM \\
\textit{Planck}+WP+Union2.1 & -1.09&0.085&1.241 & Positive for $\Lambda$CDM \\
\textit{Planck}+WP+SNLS & -1.13&0.065&0.082  & Inconclusive to weak for  $\Lambda$CDM  \\
\textit{Planck}+WP+$H_0$ & -1.24&0.090&-1.713 & Positive for $w$CDM \\
\hline
  \end{tabular}
\end{center}
\end{minipage}
\end{table*}

The Bayesian evidence for $\Lambda$CDM against a dynamical dark energy model $w$CDM with $-1.5<w_0<-0.5$ is given in the fourth column of Table \ref{Current_bayes}. Our model selection results qualitatively confirm general conclusions of the \textit{Planck} parameter estimation results \citep{Planck-Collaboration:2013aa}, where a wider prior range is used. In the \textit{Planck} paper, the BAO and Union2.1 data sets were found to be compatible with a cosmological constant, SNLS data weakly favour the phantom domain, while $H_0$ data are in tension with $w=-1$.

From our results we conclude that most of the current \textit{Planck} data favour a cosmological constant. Next we turn to the question of whether future data from the \textit{Euclid} probe can overturn these results.

\section{The PPOD applied to the dark energy question}
\label{Section_Results}

We produce a PPOD forecast for \textit{Euclid} following the method described in Section \ref{PPOD_section}. The physical question we study is the choice of dark energy model. We therefore focus on the dark energy equation-of-state parameter $w_0$, comparing a cosmological constant model ($\Lambda$CDM) with $w_0=-1$ against a dynamical dark energy model ($w$CDM) with a Gaussian prior of width $\Delta w_0=0.5$. We calculate the PPOD for two configurations of the \textit{Euclid} survey, based on the present knowledge from the 4 data sets described earlier. The constant future and current errors, $\tau$ and $\sigma$ respectively, are expressed in units of the prior width $\Delta w_0=0.5$. Thus, the current errors are $\tau= 0.0551/\Delta w_0= 0.1102$ and $\tau= 0.0382/\Delta w_0=0.0764$ for the requirement and goal survey, respectively. Likewise, we express the current mean $\mu$ and future mean $\nu$ in units of the prior width.

\subsection{PPOD forecasts}

\begin{figure}
\includegraphics[width=71.5mm,angle=270]{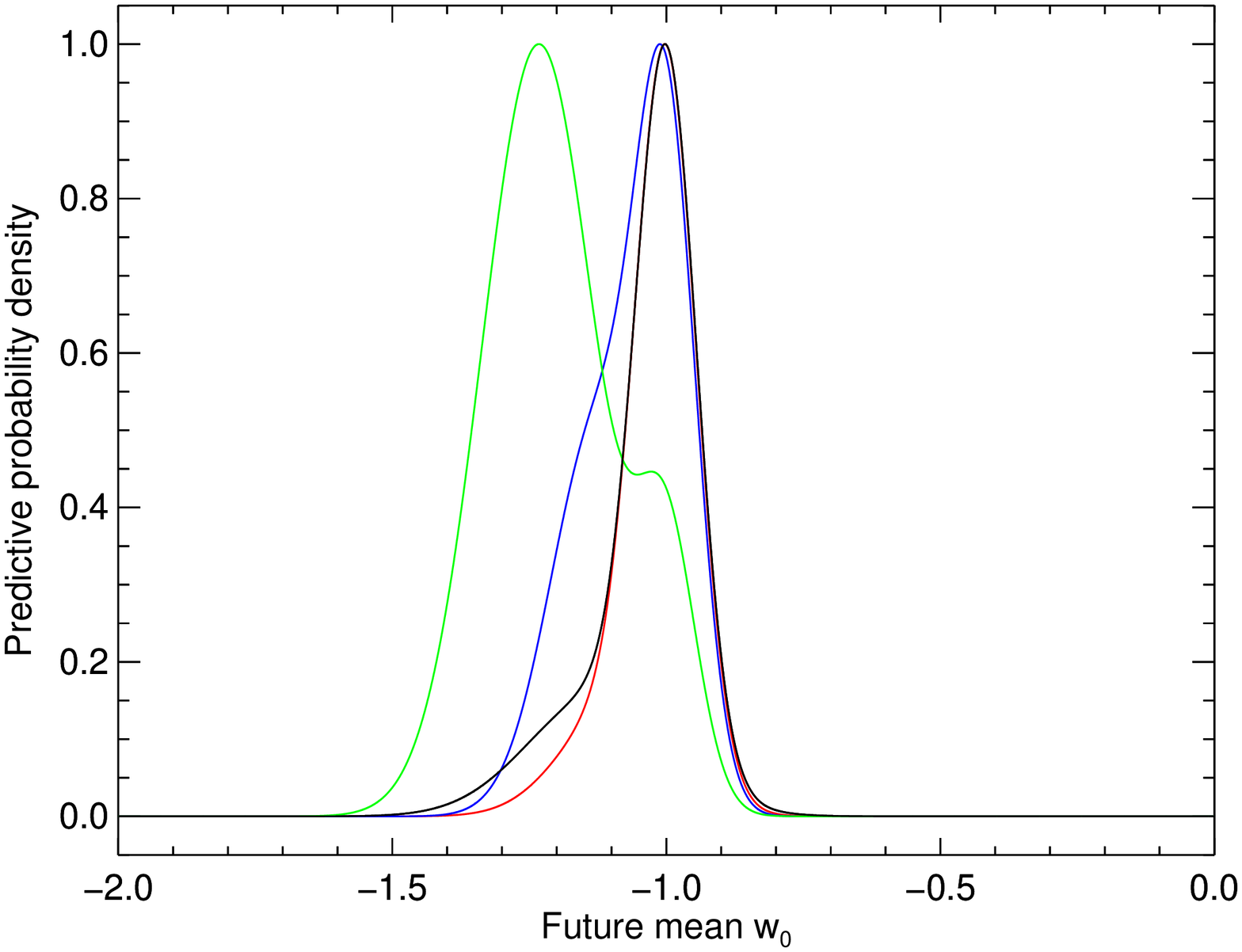}
\includegraphics[width=71.5mm,angle=270]{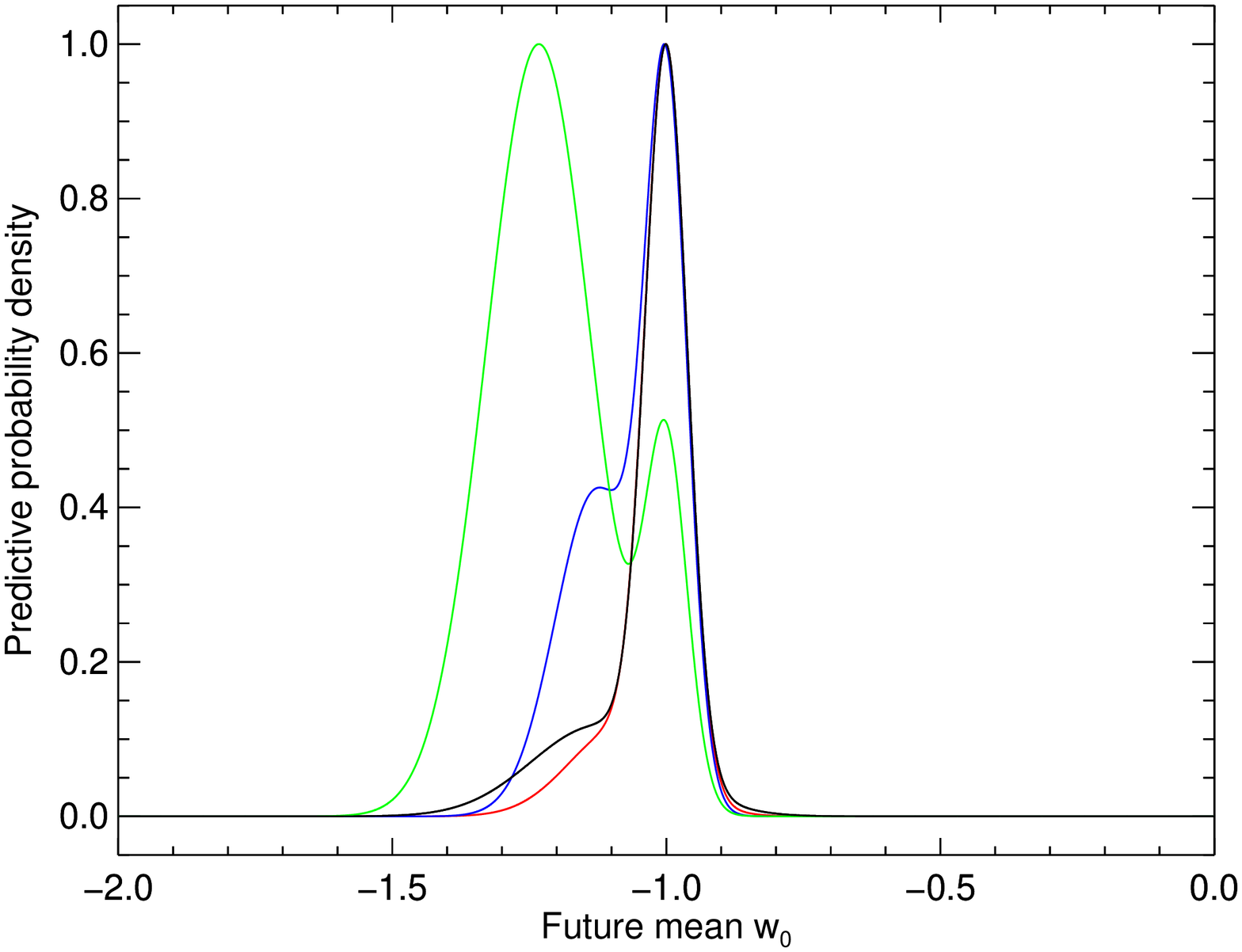}
\caption{The predictive data distribution for \textit{Euclid} cosmic shear, conditional on current knowledge. The probability distribution (normalized to the peak) is that of future measurements of the dark energy equation-of-state parameter $w_0$. The peaks centred on $w_0=-1$ correspond to $\Lambda$CDM. We use \textit{Euclid} `requirement' and `goal' survey parameters in the top and bottom panel, respectively. In each panel, we plot $p(D|d)$ for four current data sets: \textit{Planck}+WP+BAO (black), \textit{Planck}+WP+Union 2.1 (red), \textit{Planck}+WP+SNLS (blue), and \textit{Planck}+WP+$H_0$ (green).}
 \label{pdd_Euclid}
\end{figure}

The predictive distribution for \textit{Euclid} using current knowledge from \textit{Planck} is shown in Fig. \ref{pdd_Euclid}. For the \textit{Planck}+BAO+supernova data, the peak of the distribution is located at $w_0=-1$. This is a consequence of the fact that the errors around the current mean with these data sets are too large to exclude $w_0=-1$. For the \textit{Planck}+WP+$H_0$ data set, the most probable models are located around $w_0=-1.24$.

The PPOD results obtained from the predictive distribution are shown in Table \ref{Future_Bayes}. The main result is that \textit{Euclid} has a low probability of finding high-odds evidence [i.e. $p(\ln B<-5)$] for $w$CDM for the \textit{Planck} data sets using BAO and supernova data. If the current mean is given by the \textit{Planck}+WP+$H_0$ data set, then this probability increases to more than 50 per cent. This is consistent with the model selection results for \textit{Planck} given in Table \ref{Current_bayes}. It means that \textit{Euclid} is not likely to overturn the current model selection results for this choice of prior.
\begin{table*}
\begin{minipage}{180mm}
\caption{Probability of future model selection results for the two survey configurations of the \textit{Euclid} mission described in the test, using cosmic shear data, conditional on present knowledge from four \textit{Planck} data sets. The probability that \textit{Euclid} will favour $\Lambda$CDM is shown in the last column. The third to fifth columns give the probability that \textit{Euclid} will provide strong, moderate and positive evidence for $w$CDM, respectively.}
 \label{Future_Bayes}
 \begin{center}
  \begin{tabular}{@{}llcccc}
    \hline 
  \multicolumn{6}{c}{Dark energy: $w_0=-1$ versus $-1.5\le w_0 \le -0.5$ (Gaussian)}\\ 
    \hline
 Future data &Current data  & $p(\ln B < -5)$  & $p(-5< \ln B<-2.5)$ & $p(-2.5<\ln B<0)$ & $p(\ln B>0)$ \\ \hline
\textit{Euclid} requirement survey& &&&&\\
&\textit{Planck}+WP+BAO & 0.150 & 0.160 & 0.337 & 0.353 \\
&\textit{Planck}+WP+Union2.1 & 0.108 & 0.156 & 0.356 & 0.380 \\
&\textit{Planck}+WP+SNLS & 0.252 & 0.225 & 0.278 & 0.245 \\
&\textit{Planck}+WP+$H_0$ & 0.531 & 0.200 & 0.160 & 0.109 \\
\textit{Euclid} goal survey&  &&&&\\
&\textit{Planck}+WP+BAO & 0.163 & 0.135 & 0.309 & 0.393 \\
&\textit{Planck}+WP+Union2.1 & 0.125 & 0.137 & 0.323 & 0.414 \\
&\textit{Planck}+WP+SNLS & 0.286 & 0.185 & 0.253 & 0.276 \\
&\textit{Planck}+WP+$H_0$ & 0.621 & 0.119 & 0.132 & 0.128 \\
    \hline
  \end{tabular}
  \end{center}
\end{minipage}
\end{table*}

There are two points to note about the PPOD results given here. First, the Gaussian approximation used in the PPOD breaks down in the tails of the distribution. Secondly, the intervals for $\ln B$ used in the Jeffreys scale are arbitrary. Furthermore, the interpretation given to each region has an empirical origin in betting odds and depends to some extent on the nature of the model selection question \citep[see][]{Kass:1995,Efron:2001,Nesseris2013}. For these reasons, the results given here should be interpreted as a rough guide to the model selection outcome. A more general result can be obtained at the expense of computational speed by dropping the assumption of Gaussianity of the current and future likelihoods and sampling from both using Markov chain Monte Carlo techniques. 

It should be noted that the future data set considered here is the weak-lensing part of the \textit{Euclid} mission, and our results only apply to these data. \textit{Euclid} has two primary probes of dark energy, which are weak lensing and galaxy clustering. With weak lensing alone, using the goal survey parameters, we obtain a dark energy Figure of Merit (as defined in \citealt{DETF}) of $102$ \citep{Debono:2013}. The addition of galaxy-clustering data improves the Figure of Merit by a factor of $\sim4$, and would provide decisive Bayesian evidence in favour of a cosmological constant if the data are consistent with this model \citep{Laureijs:2011aa}. 

However, if there is some tension between the current maximum likelihood result and a cosmological constant, then an improvement in dark energy parameter precision will not result in a substantial improvement in the odds in favour of a cosmological constant. This is the case with the \textit{Planck}+WP+$H_0$ data. If we use the \textit{Euclid} Red Book joint weak-lensing and galaxy-clustering predicted precision of $\Delta w_0=0.015$ \citep{Laureijs:2011aa}, we obtain $p(\ln B>0)=0.142$, which is only a slight improvement on the probability of $0.128$ using the \textit{Euclid} goal survey, shown in Table \ref{Future_Bayes}, where we only consider weak-lensing future data.

On the other hand, using \textit{Planck}+WP+BAO as current data, with joint weak lensing and galaxy clustering as future data, we obtain a probability of $0.497$ for Bayesian evidence in favour of the $\Lambda$CDM model, which is a significant improvement on the lensing-only result. The point here is that an improvement in parameter precision decreases the error around the current estimated parameter value, and accumulates more evidence around the peak of the current distribution. If the current value of the peak is in tension with $\Lambda$CDM, then improved parameter precision is unlikely to overturn the current Bayesian evidence result.  For the current Bayesian evidence result to be overturned, there would have to be a systematic shift in the position of the peak of the likelihood in future data. We know this to be the case intuitively.

In our calculations for parameter errors from future data, we include statistical uncertainties, but not systematic effects, which reduce the precision and introduce bias (see e.g. \citealt{Massey2013} for a review). We expect a weak-lensing survey such as the one described here to be affected by various systematics including measurement systematics from point spread function effects \citep{KSB1995,Hoekstra2004,Mandelbaum2014}, redshift distribution systematics \citep{Hu:1999b,Ma:2006, AR2007,Cardone2014}, theoretical uncertainties on the matter power spectrum, especially in the non-linear regime \citep{Huterer:2005,Hilbert:2009,Teyssier:2009,Beynon2012}, and intrinsic correlations \citep{King:2003, Bridle2007,Joachimi:2009}.

Cosmic shear surveys are especially affected by intrinsic alignment signal contamination. A perfect knowledge of the intrinsic alignment signal would allow us to produce unbiased measurements. There is a wide variation in the impact on the degradation of the dark energy parameter errors, depending on the intrinsic alignment model \citep{Kirk2012,Heymans2013} . The dark energy Figure of Merit can be degraded by $20$--$50$ per cent depending on the intrinsic-intrinsic and galaxy-intrinsic correlation model chosen \citep{Bridle2007}.  These correlations, to a certain extent, can be mitigated by using a sufficient number of redshift bins. \citet{Laszlo2011} find that the inclusion of intrinsic alignments can change the dark energy equation-of-state Figure of Merit by a factor of $\sim 4$, but that the constraints can be recovered by combining CMB data with shear data. Provided the intrinsic alignment model is accurate enough, galaxy-intrinsic correlations themselves can be used as a cosmological probe, and the inclusion of this effect can actually enhance constraints on dark energy \citep{Kitching2011}.

Precise calculation of the errors on any parameter achievable with a weak-lensing survey is therefore highly dependent on a host of nuisance parameters and physical models. The PPOD technique only requires knowledge of the statistical properties of the parameters of interest, which means that the model selection part of the technique is independent of the parameter estimation algorithm. The parameter of interest in this work is $w_0$. The error forecasts for $w_0$ presented here are the best that can be achieved for the survey design described, using weak lensing only.

\subsection{The dependence on the prior}

The dependence of model selection conclusions on the prior range is an important aspect of modern cosmology \citep*[see e.g.][]{Kunz:2006aa}. The prior width determines the strength of the Occam's razor effect, since a larger prior favours the simpler model. While the prior range should be large enough to contain most of the likelihood volume, an arbitrarily large prior can result in an arbitrarily small evidence for the extended model. For a discussion on the dependence of evidence on the choice of prior see e.g. \citet{Kunz:2006aa}, \citet{Trotta:2007} and \citet{Brewer:2009aa}. In Section \ref{Section_Results} we used a fixed prior width of 0.5. We now examine the impact of a change of prior on our PPOD results. 

\begin{figure}
\begin{center}
\includegraphics[width=84mm]{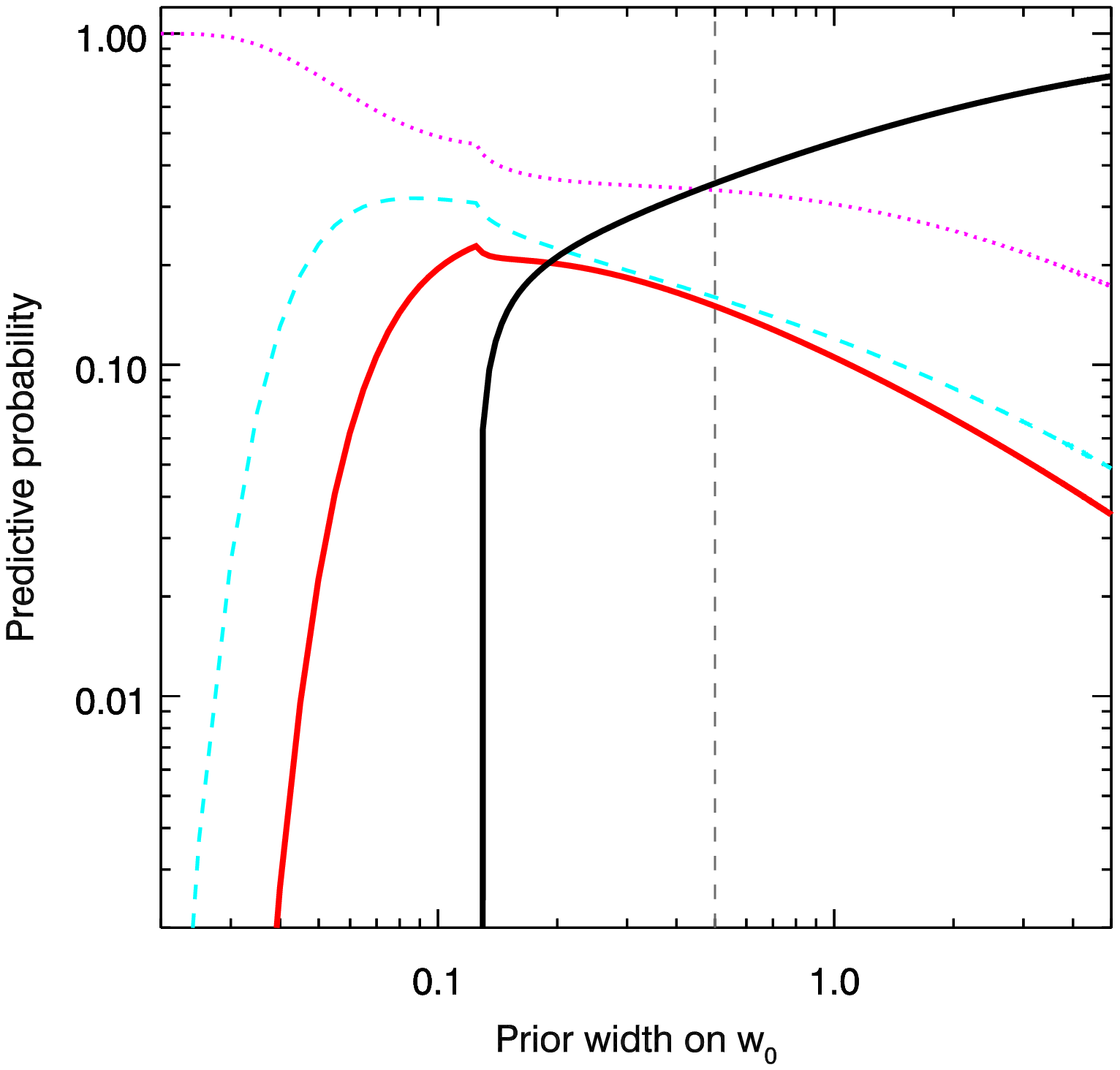}
\includegraphics[width=84mm]{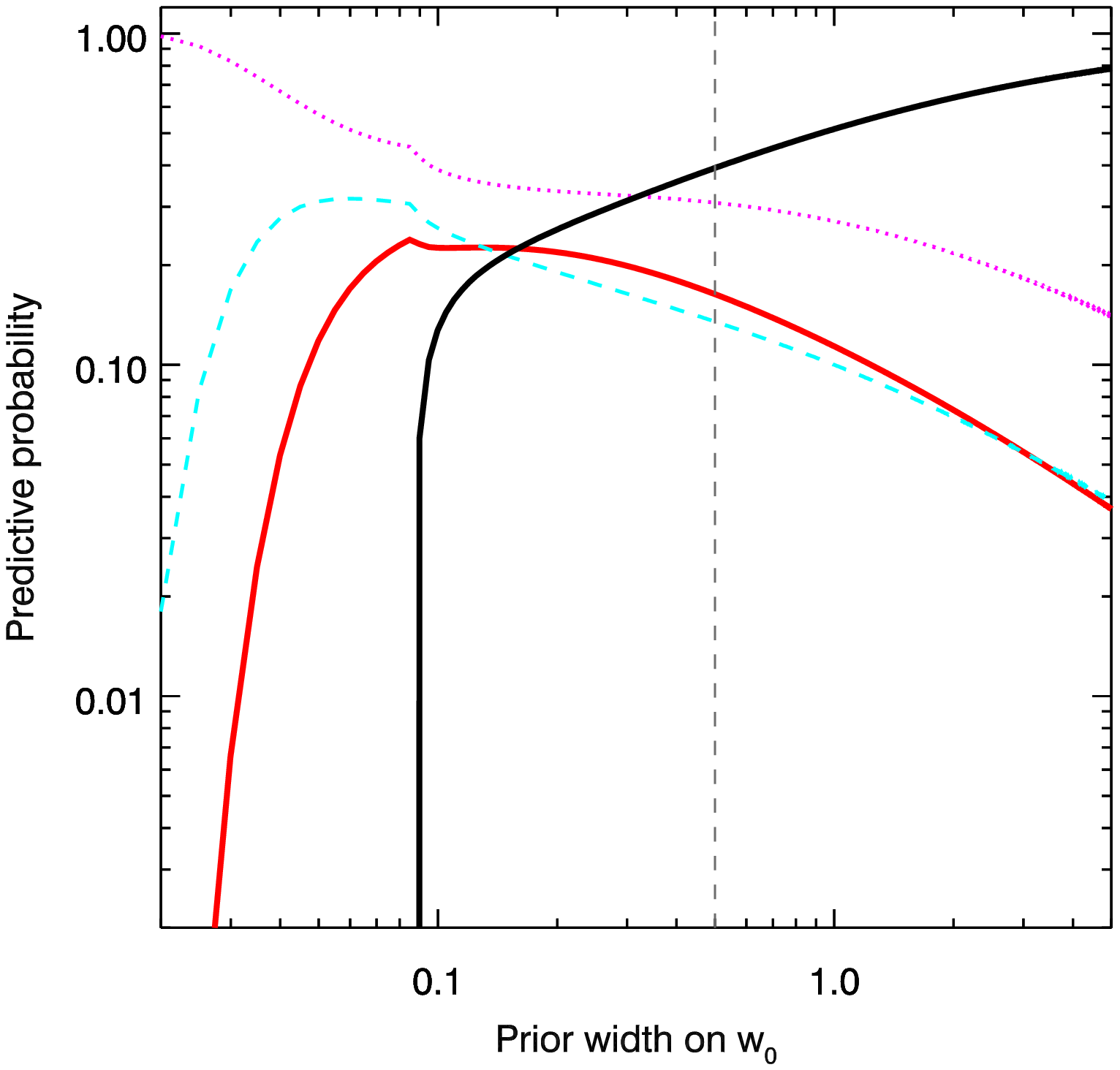}
\end{center}
\caption{The PPOD dependence on the prior width of $w_0$, using \textit{Planck}+WP+BAO as the current data, and forecasts for \textit{Euclid} with a requirement survey configuration. The vertical dotted line shows the prior width of $0.5$ used to calculate $P(D|d)$ in the previous figure. The lines show the future probability of evidence for $w$CDM according to the Jeffreys scale for Bayesian evidence: positive (magenta dots), moderate (cyan dashes), strong (thick red line) and negative (thick black line). Negative evidence for the extended model is equivalent to evidence for the restricted model $\Lambda$CDM. In order to obtain a larger probability for moderate evidence for $w$CDM one would have to use a prior width smaller than about $0.4$. With a prior width smaller than about 0.2, when the evidence for $\Lambda$CDM drops sharply, the bulk of the data will provide only positive evidence for $w$CDM, falling short of strong or even moderate evidence. In Jeffreys's terminology, the evidence is in the inconclusive regime. The probability of strong evidence for $w$CDM with this prior width is below 20 per cent. }
 \label{ppod_fig}
\end{figure}

In Fig. \ref{ppod_fig} we show the dependence on the prior width of the probabilities for \textit{Euclid} weak lensing to obtain different levels of evidence for a dynamical dark energy model $w$CDM against a cosmological constant model $\Lambda$CDM. As current data, we use the \textit{Planck}+WP+BAO data set. Our results hold for a wide range of priors. We note that the probability of evidence for $\Lambda$CDM approaches 75 per cent while the probability of strong evidence for $w$CDM falls below 10 per cent as the prior is widened beyond $\Delta w_0 = 1$. The prior would have to be narrowed to less than 0.2 for the model selection conclusion to be reversed, namely, for the probability of strong evidence for $w$CDM to be greater than the evidence for $\Lambda$CDM. For any reasonable choice of prior, and for both \textit{Euclid} survey configurations, there is at most 25 probability of strong evidence for $w$CDM.

These results show the important role of the prior in the dark energy question. The narrower the prior, the greater the precision on $w_0$ that is required to provide evidence for the extended model. The prior width defines the model predictivity space. When we seek some significant evidence for $\Lambda$CDM, we are in fact finding the space of models that can be significantly disfavoured with respect to $w_0=-1$ at a given accuracy. This point is highlighted in \citet{Trotta2006}. For an extended model with small departures from $\Lambda$CDM, it is evident that the required accuracy needs to be higher than if we were testing an extended model with large departures from $\Lambda$CDM. 

\section{Conclusions}
\label{Conclusions}

We have applied the predictive posterior odds distribution technique to produce forecasts for the Bayes factor using weak lensing from the future \textit{Euclid} probe. We carry out our calculations for the dark energy equation-of-state parameter $w_0$, which is a central parameter in the dark energy model selection question. 

We have shown that there is a high probability that cosmic shear data from \textit{Euclid} will confirm current \textit{Planck} model selection results, where the evidence for $\Lambda$CDM is positive but not overwhelming. The most important result is that for three out of four current data sets, using a prior range of $-1.5<w_0<-0.5$, future data have less than 29 per cent probability of providing strong evidence for $w_0\neq -1$. For a wider prior range, compatible with the theoretical priors of dynamical dark energy models, the probability of evidence for $\Lambda$CDM rises to above 75 per cent while the probability of strong evidence for $w_0\neq -1$ falls to less than 5 per cent. 

These conclusions qualitatively support the results in \citet{Debono:2013}, in which we find that \textit{Euclid} cosmic shear data forecasts return an undecided Bayesian evidence result if the true values of $w_0$ and $w_a$ are close to their $\Lambda$CDM fixed values of $-1$ and $0$. Furthermore, the present work shows that $\Lambda$CDM is still well supported by the forecasts if we include current information, since the inclusion of the extra parameter $w_0$ is not required by Bayesian evidence, even if the alternative model has a relatively narrow prior range ($\Delta w_0=0.5$). As we widen the prior range, the probability of evidence for $\Lambda$CDM becomes overwhelming.

Improving the parameter precision by going from a requirement to a goal-survey configuration increases the probability of evidence in favour of $\Lambda$CDM. This result holds for all prior ranges considered in this paper. The addition of galaxy-clustering data improves the parameter precision and the probability even further. Our results highlight the essential role of both parameter precision and prior width in deciding model selection questions.

\section*{Acknowledgements}
The author is supported by a European Space Agency International Research Fellowship.


\label{lastpage}

\end{document}